\documentclass[aip,apl,numerical,reprint]{revtex4-1}
\usepackage{graphicx}
\usepackage{dcolumn}
\usepackage{bm}
\usepackage{datetime}
\usepackage[version=3]{mhchem}
\usepackage{etoolbox}
\usepackage{hyperref}

\begin{document}
\title{Colorimetric path tagging of filaments using DNA-based metafluorophores}

\author{A.P. Micolich}
\affiliation{School of Physics, University of New South Wales, Sydney NSW 2052, Australia}

\date{\today}
\begin{abstract}
{\bf Summary:} The idea is a nanoscale extension of a sporting event called a `color run'.\cite{ColorRun} The filament is wrapped with a DNA origami `barrel',\cite{HwangCLEO15} which acts as the nanoscale equivalent of a white t-shirt. The DNA t-shirt has dangling ss-DNA `handles' that bind `tags' consisting of an `anti-handle' oligo bound to a fluorophore. The tagging methodology exploits the recent development of DNA-based `metafluorophores'\cite{WoehrsteinSciAdv17} with digitally tunable optical properties based on a collection of fluorophores bound to a single DNA origami structure. In this idea, the filament collects different color tags at `color-stations' as it travels through the network (Fig.~1). These contribute to a final DNA t-shirt color that represents of the path taken. Readout is implemented through lenseless on-chip imaging\cite{OzcanARBE16} via pixels in a CMOS backplane located at read-out points in the network. Color-stations are implemented via microfluidic channels on a dialysis membrane,\cite{ShengAnalyst14} which feed tags into the network in localised areas. A chemical countermeasure to prevent false tagging by stray tags that have drifted in from other stations by exploiting strand-displacement techniques\cite{ZhangNatChem11} is also proposed.
\end{abstract}

\maketitle

\noindent{\bf 1.~Requirements Summary:} Briefly summarising requirements for later reference. The goal is to encode and readout information about filament path in a network biocomputation device.\cite{NicolauPNAS16} Specific requirements are: A. ability to write information into filaments in regions a few micrometers in size; B. ability to readout at sites located hundreds of micrometers or more away from writing regions; C. writing at least $10$ ideally $100$s of tags; D. ability to tag and read at minimum $10$, ideally $100$s of different sites; E. information must be carried with the filaments; and F. readout must allow decoding of information from $>10^3$, ideally $10^6 - 10^9$ individual filaments. I will refer to these specifically as Req.~A-F in the discussion that follows. I will focus on microtubules but see no obvious reason this couldn't work for actin filaments.

\noindent{\bf 2.~Basic Concept:} A major challenge for nanoscale tagging with significant information content is space-availability. Physical barcodes scale with the minimum resolvable bit-size and the number of bits required. For Reqs.~A,C~\&~E above, a strategy to make the bit-size tiny and/or maximise encoded information-per-bit is needed. Further, rapid read-out is needed to be practical to Reqs.~D~\&~F. This provides strong incentive for solutions giving parallel or single-shot read-out of whole-path information. This inspired an idea where whole-path information is accumulated and read-out in the optical properties, e.g., color, relative brightness, photostability, of a single microscopic `pixel' (or few pixels). This colorimetric tagging would be implemented through recent advances in DNA nanotechnology, as described below. Before getting to the `nuts and bolts', I will quickly use a macroscopic analogy to illustrate the basic concept.

\begin{figure*}
\centering
\includegraphics[width = 18 cm]{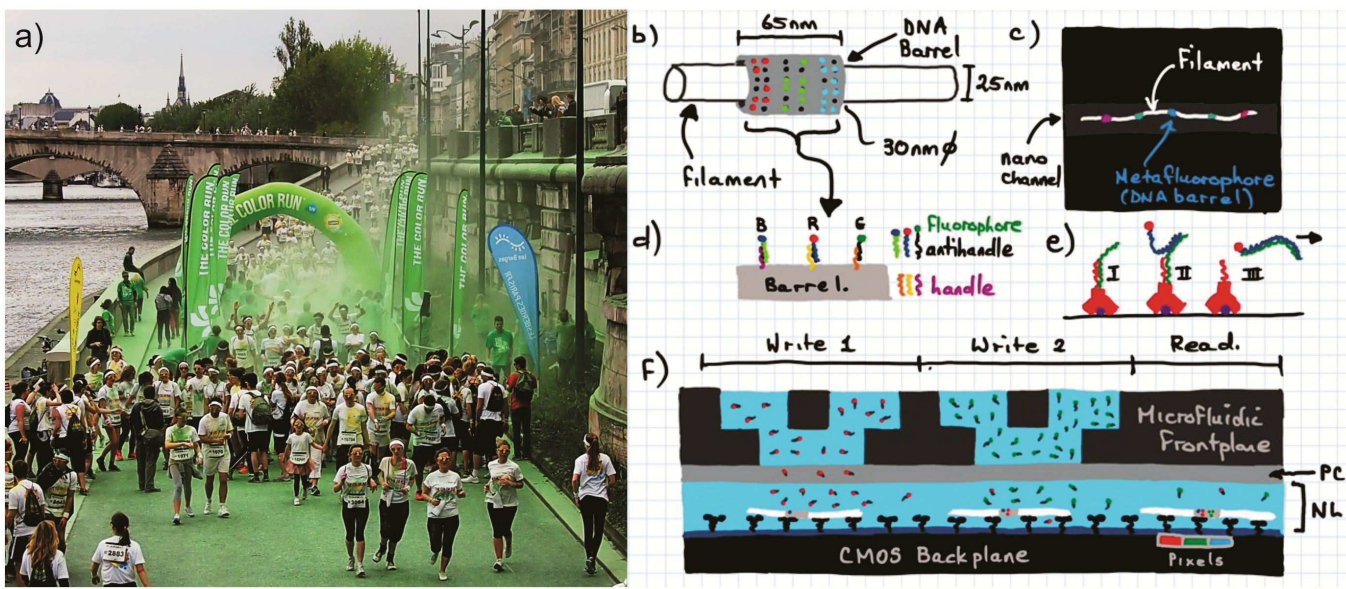}
\vspace*{1mm}
\caption{(a) Photograph illustrating basic concept (adapted from Ref.~1). (b) Diagram of DNA barrel\cite{HwangCLEO15} (t-shirt) attached to a filament. (c) Top view of filament in a flow-cell carrying multiple DNA t-shirts. (d) Schematic of the basic tagging concept.\cite{WoehrsteinSciAdv17} (e) Schematic for the chemical countermeasure for stray tags exploiting strand displacement.\cite{ZhangNatChem11} (f) Side view of a network device showing two color-stations and a read-out station. From bottom up: CMOS backplane\cite{OzcanARBE16} (black), thin glass base\cite{TakeharaAIPAdv17} (dark blue), PMMA/network flow-cell\cite{NicolauPNAS16} (light blue), dialysis membrane\cite{ShengAnalyst14} (light grey), microfluidic frontplane\cite{ShengAnalyst14, AndersonAnalChem00} (black/light blue). NL = network layer, PC = polycarbonate. Sketches not to scale.}
\end{figure*}

There is a popular sporting event called the `Color Run'.\cite{ColorRun} Participants run a $5$~km course solely for fun (no winners/no prizes). Runners start wearing a white t-shirt and run through a `color-station' at each km (Fig.~1a), where they are sprayed with a given paint colour. The run ends with colorful t-shirts and happy runners, but it also provides a means to determine if someone ran just the first km and took an Uber to the finish -- their t-shirt would be green not rainbow colored. One could easily extend this idea to a course with multiple paths, each with a station dispensing a different color and precise amount of paint. At the end, a single read-out of one parameter, t-shirt color, tells you about the path they took. To do this with filaments, we need nanoscale equivalents for: the t-shirt, paint, an ability to control the paint supply, and a way to read the color.

\noindent{\bf 3.~Usually just a t-shirt:} Microtubule-sized t-shirts don't exist yet and microtubules don't have arms, so my idea is to use a DNA origami `barrel' structure instead (Fig.~1b).\cite{HwangCLEO15, WickhamWeb, KnudsenNatNano15} These barrels can be made small enough to fit the $25$~nm microtubule diameter,\cite{HwangCLEO15} carrying up to $2000$ uniquely addressable `handles' spaced $9$~nm apart\cite{WickhamWeb} for tag-attachment. The barrels undergo two-step assembly, first forming a rectangular tile that's closed into a cylinder via staple strands that bind opposite edges of the tile. The remaining two edges can be used to polymerize barrels, with barrel trimers of $\sim100$~nm length already demonstrated.\cite{KnudsenNatNano15}

The DNA handles\cite{WickhamWeb} on the inside surface can be used to `fit' the t-shirt to the filament via an `anti-handle' oligo bound to an $\alpha$-tubulin antibody or phalloidin molecule.\cite{AgastiChemSci17} The external handles host tags (see \S4). Filaments are well-known to remain motile whilst carrying cargo, e.g., quantum dots,\cite{NitzscheNatNano08} and recent work shows no evidence of adversely-strong binding between DNA origami and kinesin\cite{WollmanNatNano14} or myosin\cite{IwakiNatComm16} that would shut-down motility.

\noindent{\bf 4.~Now add paint:} Information encoding is inspired by the DNA metafluorophores developed by Woehrstein {\it et al.}\cite{WoehrsteinSciAdv17} These feature a DNA origami tile with three `stripes' of $44$ handles. Each stripe binds anti-handles (Fig.~1d) with a different fluorophore colour (red, green or blue), giving the whole tile a highly tunable colour and intensity, hence the name `metafluorophore'. Ref.~3 used a $60~\times~90$~nm tile, which would translate to a t-shirt of circumference $80$~nm ($\approx 25\pi$~nm) and length $\sim 65$~nm (Fig~1b). The fluorophore colors are separated into stripes to reduce issues with self-quenching and FRET.\cite{WoehrsteinSciAdv17}

DNA metafluorophores have been demonstrated as single-pixel barcodes, with a maximum of $5^3~-~1~=~124$ distinct codes obtained by attributing five intensity levels to each color.\cite{WoehrsteinSciAdv17} Correct-read accuracies surpassing $95\%$ were obtained by requiring each color to have non-zero intensity (i.e., drop lowest intensity level from coding system), giving $4^3 = 64$ individual codes (see Table~S5 of Ref.~3). In Ref.~3, all $44$ handles in a stripe have the same sequence. In my idea all handles are unique\cite{WickhamWeb} and color-station specific, enabling the final color to contain information about passage via $64$ separate network points.

Disentangling color to specific path would involve a filament carrying several t-shirts with differing handle combinations. A $5~\mu$m filament could carry $15$ t-shirts and still have $80\%$ free surface for motor-binding (Fig.~1c). This could be five sets of three for redundacy/error-minimisation. One set could be pre-loaded with fluorophores to act as on-board `standard candles' during detection. The remaining four sets could display different handle combinations ($= 320$ bits available), cleverly chosen so that exact path information can be deduced from the specific color sequence that the set of t-shirts present. Devoting all $15$ t-shirts individually to this task gives $960$ distinct bits. Either case meets Req.~C. Information density could be further increased by using the photostability of different dyes of the same color, e.g., Atto-647N \& Alexa 647, as information.\cite{WoehrsteinSciAdv17}

\noindent{\bf 5. Look at the pretty lights:} Read-out is via t-shirt optical properties: color, relative intensity, possibly also photostability, etc. The $64$-bit set-up in \S4 means $18$ fluorophores minimum ($132$ max).\cite{WoehrsteinSciAdv17} This is obviously possible using traditional optical microscopy.~\cite{WoehrsteinSciAdv17} It may be a cost-effective option given super-resolution techniques can now be reasonably performed using modern cellphone CMOS imaging chips\cite{DiederichArxiv18} and this task shouldn't require super-high spatial resolution. A nicer alternative would be on-chip fluorescence microscopy.~\cite{OzcanARBE16, HuangSSC09, HongSSC17, TakeharaAIPAdv17} Huang {\it et al.}~\cite{HuangSSC09} demonstrated on-chip detection of targeted oligo-DNA (21-bp) binding using Qdot-655 fluorophores with detection limits down to $\sim 20$ fluorophores/$\mu$m$^2$ for a $50 \times 50~\mu$m pixel size in 2009. Better performance is expected now, with detection limits down to $\sim 1$ fluorophore/$\mu$m$^2$ predicted.~\cite{HuangSSC09} Both are sufficient for detection in this tagging concept. Technology for microfluidics on ultra-thin glass on CMOS for fluorescence imaging also exists.~\cite{TakeharaAIPAdv17} The underpinning CMOS imaging technology is now at single-photon detection level at room temperature,~\cite{MaOptica17} and capable of sub-micron pixels.\cite{TakahashiSensors17}

A large-area CMOS array isn't required for this idea. Instead, a custom-CMOS backplane featuring $3$-pixel (or $3n$-pixel) RGB clusters at specific read-out sites (Fig.~1f) aligned to the network structures would suffice for `single-shot' t-shirt color readout. One challenge with the t-shirts is that they are wrapped around the filament and not flat like in Ref.~3. This can be overcome by noting that microtubules rotate while gliding, typically travelling $4-10~\mu$m per $2\pi$ rotation.\cite{NitzscheNatNano08} Local enhancement by plasmonic nanoantennas could also improve detection.\cite{AcunaSci12} These nanoantennas could be nanofabricated metal bow-tie structures\cite{LeeOptEx13} built into the nanochannel floor/walls, possibly combined with gold nanoslits to mask part of the pixel for improved resolution following Gro{\ss} {\it et al.}\cite{GrossNatNano18}

\noindent{\bf 6. Time for a paint shower:} The last aspect is how to distribute antihandle-fluorophore molecules (hereafter `tags') to specific network locations. The color-station operating concept is shown in Fig.~1f. Localised addition of tags is achieved via a structured microfluidic top-plane separated from the underlying network flow chamber by a thin polycarbonate dialysis membrane.\cite{ShengAnalyst14} The membrane has nanoscale pores just large enough to allow tags to pass (Fig.~1f) but small enough to prevent filament escape.\cite{ShengAnalyst14} The membrane also provides an impedance to reduce coupling between the two fluidic systems and prevent the network flow-cell from being flooded with tags. In other words, the membrane ensures the paint is supplied as a fine spray rather than a tsunami. The membrane properties, tag concentration and flow would need to be engineered accordingly. The microfluidic top-plane would require at least two distinct PDMS layers, which is possible using conventional casting and bonding,\cite{AndersonAnalChem00} or 3D-printing methods.\cite{GlickMN16} Hence the final device structure (Fig.~1f) would be: (from bottom) CMOS backplane,\cite{TakeharaAIPAdv17} thin glass floor,\cite{TakeharaAIPAdv17} PMMA nanochannel layer,\cite{NicolauPNAS16} polycarbonate membrane,\cite{ShengAnalyst14} and microfluidics layers.\cite{AndersonAnalChem00}

\noindent{\bf 7. Blowing in the wind:} An issue in this approach is tags drifting away from their station, giving false signal. My idea includes a chemical countermeasure for deactivating stray tags (Fig.~1e). The nanochannel floors are coated in either casein\cite{NicolauPNAS16} or biotin.\cite{KeyaSciRep17} Short ss-DNA oligos are bound to either anti-casein or streptavidin and primed to the nanochannel surfaces via the color-stations prior to starting computation. I will refer to these as `floor-strands' (red in Fig.~1e). A series of `deactivation handles' (green in Fig.~1e) are made for each color-station; these are partially complementary to the local floor strands and fully complementary to tags from adjacent/upstream color-stations. Note that these also need to be orthogonal (at least very weakly complementary) for their own color-station's tags. The deactivation handles are flushed through prior to computation to prime the floor-strands (Fig.~1e-I) and supplied continuously at low flow during computation to replenish deactivation handles lost to the following deactivation process. The deactivation handles function by capturing stray tags (blue with red spot in Fig.~1e-II) from adjacent color-stations. In the process the deactivation handle detaches from the floor-strand by `strand displacement',\cite{ZhangNatChem11} with the deactivated stray tag leaving the flow channel as waste in the buffer flow (Fig.~1e-III). The floor-strand is then ready to capture a new deactivation handle from its color-station, returning to the state in Fig.~1e-I to await another stray tag. Noting that the network flow-cell is operated under directional flow for ATP supply, deactivation should only be required for upstream color-stations. Misdirected handle-tag hybridization may be further minimised by careful handle sequence choice, especially between adjacent color-stations.

\noindent{\bf Acknowledgement:} I thank Lawrence Lee for many fun discussions on DNA origami and Anders Kyrsting for introducing me to lenseless imaging.

\end{document}